\documentclass[prl,11pt,preprint]{revtex4-1}
\usepackage{graphics}
\usepackage[mathcal]{euscript}
\usepackage[latin1]{inputenc}
\usepackage{latexsym}
\usepackage{amsmath}
\usepackage{hyperref}
\usepackage{graphicx}   
\usepackage{verbatim}  
\usepackage{color}      
\usepackage{subfigure}  
\usepackage{bm}
\usepackage{amssymb}
\usepackage{bbm}
\usepackage{float}
\usepackage{slashed}
\usepackage{amsfonts}
\usepackage{euscript}   
\usepackage{mathrsfs} 
\usepackage{bbding}
\usepackage{cancel}
\usepackage[usenames,dvipsnames]{xcolor}
\hypersetup{colorlinks=true, citecolor=Plum, linkcolor=Plum,
urlcolor=Plum}

\begin{document}

\title{Stellar equilibrium in semiclassical gravity}

\author{Ra\'ul Carballo-Rubio}
\email{raul.carballorubio@sissa.it}
\affiliation{SISSA, International School for Advanced Studies, Via Bonomea 265, 34136 Trieste, Italy}
\affiliation{INFN Sezione di Trieste, Via Valerio 2, 34127 Trieste, Italy}            
\affiliation{Department of Mathematics \& Applied Mathematics, University of Cape Town, 7701 Rondebosch, South Africa}

\bigskip
\begin{abstract}

The phenomenon of quantum vacuum polarization in the presence of a gravitational field is well understood and is expected to have a physical reality, but studies of its back-reaction on the dynamics of spacetime are practically non-existent outside the specific context of homogeneous cosmologies. Building on previous results of quantum field theory in curved spacetimes, in this letter we first derive the semiclassical equations of stellar equilibrium in the $s$-wave Polyakov approximation. It is highlighted that incorporating the polarization of the quantum vacuum leads to a generalization of the classical Tolman-Oppenheimer-Volkoff equation. Despite the complexity of the resulting field equations, it is possible to find exact solutions. Aside from being the first known exact solutions that describe relativistic stars including the non-perturbative backreaction of semiclassical effects, these are identified as a non-trivial combination of the black star and gravastar proposals.

\end{abstract}

\maketitle

\def\HRULE{{\bigskip\hrule\bigskip}}


\noindent
\textsl{Introduction.--}The recent detection of gravitational waves \cite{Abbott2016} has revived interest (e.g., \cite{Cardoso2016a,Cardoso2016,Abedi2016,Barcelo2017,Price2017,Nakano2017,Volkel2017,Mark2017,Maselli2017,Krishnendu2017} and references therein) in theoretical scenarios in which black holes are replaced by horizonless ultra-compact configurations. Current electromagnetic and gravitational wave data definitively leave room for these alternatives to exist \cite{Cardoso2017n,Cardoso2017}. This letter does not discuss the diverse motivations for these theoretical considerations or review different proposals available in the literature, for which we refer the reader to \cite{Cardoso2017,Visser2009,Barcelo2015u} for instance; it focuses rather on the following question: Is it possible to obtain horizonless ultra-compact configurations from known physics?

The theoretical status of horizonless ultra-compact objects is far from satisfactory at best, as their construction has been so far based on assuming large deviations from known physics, postulating geometries with no complete mathematical framework to justify the origin of these deviations, or to control that other predictions of general relativity are not affected. The lack of theoretical constraints leads to huge uncertainties on the properties of these hypothetical objects, even if considering the simpler problem of describing static configurations.

This letter aims to represent a first step in changing this situation, providing a framework in which the theoretical assumptions are minimal and under control, and everything else is obtained just by solving a consistent set of field equations. In particular, here we deal with the most conservative extension of general relativity, the semiclassical Einstein field equations \cite{Hartle1981,Flanagan1996,Hu2001}
\begin{equation}
G_{\mu\nu}=8\pi G(T_{\mu\nu}+\hbar N\langle\hat{T}_{\mu\nu}\rangle)+\mathscr{H}_{\mu\nu},
\label{eq:sefq}
\end{equation}
using a standard definition of the renormalized stress-energy tensor $\langle\hat{T}_{\mu\nu}\rangle$ describing the quantum vacuum polarization of $N\gg 1$ matter fields. $\mathscr{H}_{\mu\nu}$ stands for terms that, while being generally non-zero, are negligible for the situations studied in this letter. These can be further classified into two categories. The first category contains $\mathscr{O}(1/N)$ contributions with respect to the term proportional to $\hbar\langle\hat{T}_{\mu\nu}\rangle$ (i.e., subleading terms in the $1/N$ expansion). The second category includes contributions from possible higher-derivative curvature corrections to the Einstein-Hilbert Lagrangian which, on general grounds, can be written schematically as $\hbar^{n-1}G^{n-2}\mathscr{R}^n$ with $n\geq2$ and $\mathscr{R}^n$ a scalar polynomial of order $n$ on the Riemann curvature or its derived curvature tensors. For instance, the $n=2$ contributions are needed in order to renormalize $\langle\hat{T}_{\mu\nu}\rangle$ \cite{Birrell1982,Visser2002} and are proportional to $N$. The semiclassical approximation is meaningful as long as curvature remains small enough. Hence, in practice we will study the field equations \eqref{eq:sefq} with $\mathscr{H}_{\mu\nu}=0$ identically, justifying later that this is indeed a good approximation for the solutions to be described.

We start presenting the new set of equations of stellar equilibrium that follows from these semiclassical equations, and continue with the description of a family of exact solutions. Let us remark that here we deal with static situations only. Our results do not imply by themselves that (perhaps some) astrophysical black holes are horizonless ultra-compact objects, as this conclusion could only follow from a detailed analysis of dynamical situations. However, that it is possible to obtain the static properties of such hypothetical configurations from a set of solid principles certainly makes this possibility more plausible from a theoretical perspective, and is therefore of clear relevance for dynamical studies.


\noindent
\textsl{Setting.--}Our first goal is showing that the semiclassical Einstein field equations \eqref{eq:sefq} can be expressed in a closed and simple way under the two following mild assumptions: (i) spherical symmetry: this is common in the analytical study of relativistic stars in hydrostatic equilibrium. For instance, the Tolman-Oppenheimer-Volkoff (TOV) equation \cite{Tolman1930,Oppenheimer1939}, to be generalized here, is found in classical general relativity in this approximation; (ii) $s$-wave Polyakov approximation: this neglects quantum fluctuations that are not spherically symmetric, as well as the effects of backscattering, by means of a dimensional reduction (projection) to a 2-dimensional manifold. This is a well-known approximation that is routinely used in black hole physics and permits us to obtain closed analytical expressions for quantities of interest in the semiclassical theory (e.g., \cite{Birrell1982,Nojiri1999,Fabbri2005book}), preserving the qualitative features of all the relevant vacuum states \cite{Davies1976,Visser1996}.

Hence we will deal with spherically symmetric spacetimes, with line element
\begin{equation}
\text{d}s^2=\text{d}s^2_{(2)}+r^2\text{d}\Omega^2=g_{ab}(y)\text{d}y^a\text{d}y^b+r^2(y)\text{d}\Omega^2(\theta,\varphi),\label{eq:sph4}
\end{equation}
where $\text{d}\Omega^2(\theta,\varphi)$ is the angular line element on the 2-sphere. Greek indices such as $\mu$, $\nu$, take four values, while latin indices such as $a$, $b$, take just two values. For static situations, the metric in Eq. \eqref{eq:sph4} has a time-like Killing vector field $\xi$.

Regarding the classical source in Eq. \eqref{eq:sefq}, for simplicity we consider a perfect fluid in a spherically symmetric and static configuration. The stress-energy tensor is therefore given by $T_{\mu\nu}=(\rho+p)u_\mu u_\nu+pg_{\mu\nu}$, where the velocity field of the fluid $u$ is the vector field associated with the time-like Killing vector field $\xi$ of the spacetime geometry \eqref{eq:sph4} but properly normalized such that $u_\mu u^\mu=-1$. This guarantees that we are describing a static configuration.

To this classical source, we are adding the vacuum polarization as described by the (renormalized) vacuum expectation value of the stress-energy tensor of $N$ non-interacting scalar fields (following the usual practice \cite{Hartle1981,Flanagan1996,Hu2001}). This additional piece is given in the $s$-wave Polyakov approximation \cite{Fabbri2005book} by $N$ times
\begin{equation}
\langle\hat{T}_{\mu\nu}\rangle=\frac{\delta^a_\mu\delta_\nu^b}{4\pi r^2}\langle\hat{T}_{ab}\rangle^{(2)},\label{eq:semisource}
\end{equation}
where $\langle\hat{T}_{ab}\rangle^{(2)}=\langle0|\hat{T}_{ab}|0\rangle^{(2)}$ is evaluated in the 2-dimensional spacetime $\text{d}s^2_{(2)}=g_{ab}(y)\text{d}y^a\text{d}y^b$ \cite{Polyakov1981,Maggiore2016}. In order to describe a static configuration, the expectation value in Eq. \eqref{eq:semisource} is taken in the Boulware state $|0\rangle$, namely the state associated with the time-like Killing vector field $\xi$.


\noindent
\textsl{Quantum vacuum polarization in 2 dimensions.--}Let us recall how to calculate $\langle\hat{T}_{ab}\rangle^{(2)}$ in the Boulware state \cite{Davies1977,Davies1977b}. For spherically symmetric spacetimes, the 2-dimensional line element in Eq. \eqref{eq:sph4} can be always put in the form
\begin{equation}
\text{d}s^2_{(2)}=-C(r)\text{d}t^2+\frac{\text{d}r^2}{1-2Gm(r)/r}.\label{eq:sph2}
\end{equation}
A closed expression for the renormalized stress-energy tensor $\langle\hat{T}_{ab}\rangle^{(2)}$ is most easily obtained in null coordinates, and then it is transformed back to the initial $(t,r)$ coordinates. Equivalently, the renormalized stress-energy tensor can be written in an explicitly tensorial way \cite{Barcelo2011} as $\langle\hat{T}_{ab}\rangle^{(2)}=\left(R^{(2)}g_{ab}+A_{ab}-\frac{1}{2}g_{ab}A\right)/48\pi$, where $R^{(2)}$ is the Ricci scalar of the 2-dimensional metric with line element \eqref{eq:sph2} and $A_{ab}=4|\xi|^{-1}\nabla_a\nabla_b|\xi|$ with $\xi=\partial_t$ for the Boulware state, so that $|\xi|=\sqrt{C(r)}$. The result of these computations are the following components of the renormalized stress-energy tensor in the Boulware state:
\begin{align}
24\pi\langle\hat{T}_{rr}\rangle^{(2)}=&-\frac{1}{4}\left(\frac{C'}{C}\right)^2,\qquad \langle\hat{T}_{tr}\rangle^{(2)}=\langle\hat{T}_{rt}\rangle^{(2)}=0,\nonumber\\
24\pi\langle\hat{T}_{tt}\rangle^{(2)}=&\left(1-\frac{2Gm}{r}\right)C''-C'\left(\frac{Gm}{r}\right)'\nonumber\\
&-\frac{3}{4}\left(1-\frac{2Gm}{r}\right)\frac{{C'}^2}{C}.\label{eq:2renc}
\end{align}
From now on, $f'=\text{d}f/\text{d}r$ for any function $f$.


\noindent
\textsl{Field equations and semiclassical TOV equation.--}The semiclassical source in Eq. \eqref{eq:semisource} is identically conserved, as it can be checked explicitly using its components in Eq. \eqref{eq:2renc}. The Bianchi identities imply then the conservation of the classical source $T_{\mu\nu}$, which translates into the usual continuity equation
\begin{equation}
p'=-\frac{1}{2}(\rho+p)\frac{C'}{C}.\label{eq:conteq}
\end{equation}
Taking into account that the Einstein tensor $G_{\mu\nu}$ of the geometry \eqref{eq:sph4} and both classical and semiclassical sources are diagonal, this amounts in principle to five differential equations: the diagonal components $(t,t)$, $(r,r)$, $(\theta,\theta)$, and $(\varphi,\varphi)$ of the semiclassical field equations \eqref{eq:sefq}, plus the continuity equation \eqref{eq:conteq} above. However, as in general relativity, only three of these five equations are independent. We choose to work with the continuity equation \eqref{eq:conteq} and the $(t,t)$ and $(r,r)$ components of the semiclassical field equations \eqref{eq:sefq}. These two are given, respectively, by
\begin{align}
\frac{2Gm'}{r^2}=\,&8\pi G \rho+\frac{\ell_{\rm P}^2}{r^2}\left[\left(1-\frac{2Gm}{r}\right)\frac{C''}{C}\right.\nonumber\\
&-\frac{C'}{C}\left(\frac{Gm}{r}\right)'-\left.\frac{3}{4}\left(1-\frac{2Gm}{r}\right)\left(\frac{C'}{C}\right)^2\right],\label{eq:semitemp}
\end{align}
(where we have defined the ``renormalized'' Planck length as $\ell_{\rm P}^2=\hbar G N/12\pi$), and
\begin{equation}
\frac{C'}{rC}-\frac{2Gm}{r^2(r-2Gm)}=\frac{8\pi Gp}{\displaystyle1-2Gm/r}-\frac{\ell_{\rm P}^2}{4}\left(\frac{C'}{rC}\right)^2.\label{eq:radeq}
\end{equation}
It is illuminating to combine the latter equation with the continuity equation \eqref{eq:conteq} to obtain the semiclassical TOV equation
\begin{align}
p'\left(1-\frac{\ell_{\rm P}^2}{2r}\frac{p'}{\rho+p}\right)=-\frac{Gm}{r^2}\rho\frac{(1+p/\rho)(1+4\pi r^3 p/m)}{1-2Gm/r}.\label{eq:semitov}
\end{align}
The right-hand side of this equation is the same as the right-hand side of the well-known classical equation, and it is arranged in a way that relativistic modifications beyond Newtonian gravity are easily identified. The left-hand side contains new (dimensionless) contributions describing semiclassical modifications due to vacuum polarization. Eq. \eqref{eq:semitov} is then an extension of the TOV equation that describes the hydrostatic equilibrium of (spherically symmetric) relativistic stars including semiclassical effects. The equations above can be written more compactly in terms of the enthalpy $h(r)$, defined through the relation $h'/h=p'/(\rho+p)$.


\noindent
\textsl{Exact solutions.--}The semiclassical TOV equation \eqref{eq:semitov} is a second-order polynomial equation for the gradient of the pressure $p'$, which has two roots. One of these two roots leads to the classical TOV equation in the formal limit $\hbar\rightarrow0$. The other root describes solutions for which the semiclassical modifications in Eq. \eqref{eq:semitov} are dominant, and are therefore non-perturbative (i.e., cannot be obtained by a perturbative expansion around classical solutions). The new exact solutions we have found are associated with this second root, 
\begin{equation}
p'=\frac{r(\rho+p)}{\ell_{\rm P}^2}\left(1+\sqrt{\displaystyle1+\frac{2G\ell_{\rm P}^2}{r^3}\frac{m+4\pi r^3 p}{1-2Gm/r}}\right).\label{eq:posroot}
\end{equation}
As in the classical theory, an additional constraint (that usually takes the form of an equation of state relating directly $p$ and $\rho$) is needed in order to guarantee that there are the same number of differential equations than independent unknown functions. The constraint that the quantity inside the square root in the previous equation equals a constant $\lambda^2$, with $\lambda>1$, permits us to solve exactly the previous equation and obtain
\begin{align}
&m(r)=\frac{r}{2G}\left[1+\frac{f(r)}{\lambda-1}\right]\frac{1}{1+\ell_{\rm P}^2/r^2(\lambda^2-1)},\nonumber\\
&\rho(r)=\frac{(1+\lambda)f(r)}{8\pi G\ell_{\rm P}^2}-\frac{f'(r)}{8\pi G r},\nonumber\\
&p(r)=-\frac{(1+\lambda)f(r)}{8\pi G\ell_{\rm P}^2}.\label{eq:soll1}
\end{align}
The details of this integration are not needed for our discussion here, as the assertion that these expressions solve Eq. \eqref{eq:posroot} and satisfy the mentioned constraint can be confirmed by direct substitution. On the other hand, using Eq. \eqref{eq:soll1}, Eq. \eqref{eq:conteq} can be now integrated to yield
\begin{equation}
C(r)=C(R)e^{(1+\lambda)(R^2-r^2)/\ell_{\rm P}^2}.\label{eq:soll4}
\end{equation}
The value of the integration constant $C(R)$ will be determined below. The function $f(r)$ in Eq. \eqref{eq:soll1} satisfies $f(R)=0$ in order to ensure the standard condition that the pressure vanishes at the surface of the star, $p(R)=0$.

Eqs. \eqref{eq:soll1} and \eqref{eq:soll4} solve identically Eqs. \eqref{eq:conteq} and \eqref{eq:posroot}, and therefore Eqs. \eqref{eq:radeq} and \eqref{eq:semitov}. The expressions we have obtained display an unknown function $f(r)$. If these expressions are inserted into the only equation that remains to be solved, namely Eq. \eqref{eq:semitemp}, it results in a differential equation for $f(r)$ that can be integrated. Hence, we have reduced the problem (at least for this family of solutions) to the integration of a first-order ordinary differential equation that has a unique solution satisfying the boundary condition $f(R)=0$ for each value of $\lambda$. It is straightforward to obtain this differential equation, as well as analytical expressions for $f(r)$ for different values of $\lambda$. It is more useful for our purposes here to note that its leading order in $(\ell_{\rm P}/r)^2$ is given by
\begin{align}
f(r)=\frac{\ell_{\rm P}^2(1+\mathscr{O}[(\ell_{\rm P}/r)^2])}{(1+\lambda)r^2}\left[1-\frac{r^2}{R^2}\,e^{(1+\lambda)(r^2-R^2)/\ell_{\rm P}^2}\right].\label{eq:ssol5}
\end{align}
Eq. \eqref{eq:ssol5} permits to obtain simple expressions at leading order in $(\ell_{\rm P}/r)^2$ for all the physical quantities, e.g., pressure and density:
\begin{align}
&p(r)=\frac{-1+\mathscr{O}[(\ell_{\rm P}/r)^2]}{8\pi G r^2R^2}\left[R^2-r^2e^{(1+\lambda)(r^2-R^2)/\ell_{\rm P}^2}\right],\nonumber\\
&\rho(r)=\frac{1+\mathscr{O}[(\ell_{\rm P}/r)^2]}{8\pi G r^2R^2}\left[R^2+r^2e^{(1+\lambda)(r^2-R^2)/\ell_{\rm P}^2}\right].\label{eq:presdens}
\end{align}
The three point-wise energy conditions \cite{Poisson2004}, null, weak, and dominant are satisfied as $\rho(r)+p(r)>0$, but the strong energy condition would be violated by the perfect fluid alone. As the strong energy condition must apply to the complete source of the field equations (it is ultimately a statement about the Ricci tensor), that it is violated by the classical source only does not have any physical significance, as this very configuration of the perfect fluid cannot exist by itself: the semiclassical source must be included. The three energy conditions that are satisfied have a clear physical meaning in terms of the local properties of the perfect fluid, and guarantee that its properties are that of standard matter (positive density and causal fluxes along time-like and null trajectories). Note that the perfect fluid reaches densities that are many orders of magnitude greater than the typical average density of a neutron star.


\noindent
\textsl{Validity of the approximation.--}The renormalized stress-energy tensor on the right-hand side of Eq. \eqref{eq:sefq} is the leading order in the $1/N$ expansion of quantum matter fields \cite{Hartle1981,Flanagan1996,Hu2001}. For the solutions described above this semiclassical contribution is comparable to the classical contribution $T_{\mu\nu}$. It is therefore necessary to show that higher orders are negligible, namely that these solutions are consistent with the truncation $\mathscr{H}_{\mu\nu}=0$ of the field equations \eqref{eq:sefq}. In static situations (to which this letter is restricted) it is enough to realize that (i) terms in the first category of $\mathscr{H}_{\mu\nu}$ are suppressed at least as $1/N$, and (ii) curvature is well below Planckian values almost everywhere for these solutions, which makes non-suppressed higher-derivative curvature terms irrelevant. The first consideration is a corollary of the $1/N$ expansion, but the second one is non-trivial, as not all possible solutions of Eq. \eqref{eq:sefq} have to verify it necessarily. Using the expressions above for the solutions described, the Ricci scalar and other curvature invariants can be calculated and shown to be small (in units of $\ell_{\rm P}^{-2}$) for $r/\ell_{\rm P}\gg 1$. Curvature invariants become $\mathscr{O}(\ell_{\rm P}^{-2})$ only in the region in which $r\leq\mathscr{O}(\ell_{\rm P})$, which for typical values of $R$ and $N$ is many (more than 30) orders of magnitude smaller than the radius $R$ of the star; the semiclassical (as well as the classical) field equations are no longer reliable in this region, which is reasonable as quantum gravity is expected to be necessary in order to describe these small distances. The expressions above display a curvature singularity in the $r\rightarrow0$ limit, which is however inside this region in which the semiclassical approximation (and therefore these expressions) cannot be trusted. Pending a detailed analysis of the quantization of the geometry of the exact solutions described in this letter, this issue can be approached from an effective perspective as it is done in the study of regular black holes (e.g., \cite{Hayward2005}), for instance, replacing $r^2\rightarrow r^2+\ell_{\rm P}^2$ and $R^2\rightarrow R^2+\ell_{\rm P}^2$ in Eq. \eqref{eq:ssol5}. This prescription leads to regular geometries and guarantees that all physical quantities remain bounded.
 

\noindent
\textsl{Regarding stability.--} Potential alternatives to black holes should be stable (or decay over long enough timescales). A first non-trivial check of the solutions studied here can be motivated by analogy with the analysis of secular stability in the classical theory (see the Chap. 9 in \cite{Friedman2013book} and references therein). The density at the center of the structure is given by $\rho_{\rm c}=\left.m'(r)/4\pi r^2\right|_{r=0}$, while from Eq. \eqref{eq:soll1} it follows that $M=m(R)=R/2G[1+\ell_{\rm P}^2/R^2(\lambda^2-1)]$ is a monotonically increasing function with $R$. Calculating the derivative of these functions along the parameter of the family $\lambda$, it is straightforward to show that $\text{d}M/\text{d}\rho_{\rm c}=4\pi\ell_{\rm P}^4\{1+\mathscr{O}[(\ell_{\rm P}/R)^2]\}/3R(\lambda^2-1)^2>0$ for all the solutions in the family, so that there are no turning points. The occurrence of turning points would have pointed to the unstable nature of the bulk by itself. The word ``bulk'' is used here in order to emphasize that, as described just below, these exact solutions display non-trivial surface properties. Hence stability cannot be concluded without taking into account these additional features of the surface and their interplay with the bulk dynamics. This analysis, which is out of the scope of this letter, would follow similar conceptual steps as in the gravastar proposal \cite{Visser2003,Pani2009}; note however that the junction conditions are different than in general relativity. An additional subtlety is associated with the regularization of the core with size roughly of $\ell_{\rm P}$, which implies that suitable boundary conditions must be imposed on perturbations at the boundary of the core.

\begin{figure}[H]%
\begin{center}
\includegraphics[width=0.4\textwidth]{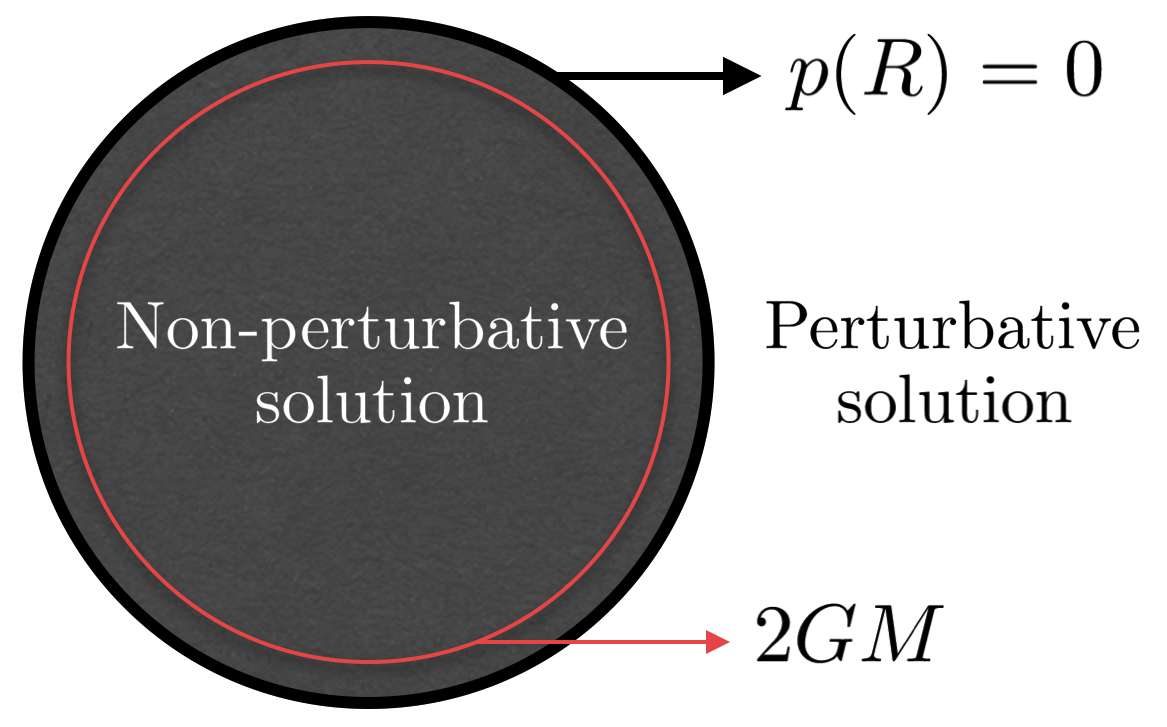}
\end{center}
\caption{The complete geometry of static ultra-compact stars in semiclassical gravity. The exterior geometry (white) is a perturbative solution of the semiclassical Einstein field equations, while the interior geometry (dark gray) is a non-perturbative solution. Interior and exterior geometries are glued at the surface of the star, the position of which is determined by the condition that the pressure of the perfect fluid vanishes. The difference between $R$ and $2GM$ has been vastly exaggerated for illustrative purposes.}
\label{fig:fig1}%
\end{figure}

\noindent
\textsl{Junction conditions at the surface.--}As in general relativity, the exact interior solution describing the geometry inside the perfect fluid has to be glued with the exterior geometry outside the fluid at the surface in which the pressure vanishes, $p(R)=0$ (see Fig. \ref{fig:fig1}). The surface is located at $r=R=2GM+\gamma\ell_{\rm P}^2/2GM>2GM$ with a minimum value $\gamma=\mathscr{O}(1)$. The numerical integration of the semiclassical field equations for the Boulware state outside the gravitational radius is well understood and has been carried out for instance in \cite{Fabbri2005,Fabbri2005b,Ho2017}, resulting that the exterior geometry is approximated well by the Schwarzschild geometry but close to the Schwarzschild radius. Using these results permits us to fix the two unspecified constants as $C(R)=1-2GM/R+\mathscr{O}[(\ell_{\rm P}/R)^2]>0$ and  $\lambda-1=\mathscr{O}(1)>0$, where the particular value of the latter is tied up to the value of $\gamma$. 

The root of Eq. \eqref{eq:semitov} that must be used in order to obtain the correct classical limit far away from the relativistic star verifies $\lim_{r\rightarrow R^+}\left.C'/C\right|_{+}=-2 R(1-\lambda)/\ell_{\rm P}^2$, while for the interior solution we have from Eqs. \eqref{eq:conteq} and \eqref{eq:posroot} that $\lim_{r\rightarrow R^-}\left.C'/C\right|_{-}=-2R(1+\lambda)/\ell_{\rm P}^2$. This implies that the surface of the star has a non-zero surface tension; i.e., it behaves similarly to a soap bubble. Writing $C'/C=\left.C'/C\right|_{+}\theta(r-R)+\left.C'/C\right|_{-}\theta(R-r)$ with the usual convention $\theta(0)=1/2$, it is possible to show that the semiclassical field equations are well-defined in the distributional sense (following a similar analysis as in general relativity, e.g., \cite{Poisson2004}). The corresponding distributional source at $r=R$ has a stress-energy tensor $S_{\mu\nu}\delta(r-R)\sqrt{1-2M/R}$ \cite{Mansouri1996}. From Eq. \eqref{eq:semitemp} it follows that the surface energy density is $S_{00}\sqrt{1-2M/R}=\Sigma=-\lambda\ell_{\rm P}^2/2\pi G R^3[(\lambda^2-1)+\ell_{\rm P}^2/R^2]$. This represents a negative but small surface density with respect to the value of the bulk density at the surface, $|\Sigma/R\rho(R)\sqrt{1-2M/R}|=\mathscr{O}\left(\ell_{\rm P}/R\right)\ll1$. On the other hand, the angular equations, that we have not discussed explicitly above (as these follow from the conservation equation), contain terms proportional to $C''/C$. The associated surface tension is $S_{\varphi\varphi}\sqrt{1-2M/R}=\sin^2\theta S_{\theta\theta}\sqrt{1-2M/R}=\sin^2\theta\,\Pi$, with $\Pi=\lambda/4\pi GR[(\lambda^2-1)+\ell_{\rm P}^2/R^2]$. The radial pressure cannot have a distributional part, as all the remaining quantities in Eq. \eqref{eq:radeq} are continuous. Hence an alternative way to obtain $\Pi$, and which needs only the independent equations used for the analysis in the bulk, is extending Eq. \eqref{eq:conteq} to include an anisotropic pressure $p_{\theta}=p+\Pi\delta(r-R)$ and density $\rho+\Sigma\delta(r-R)$, both containing a distributional part. This extension is given (see \cite{Lemaitre1933,Cattoen2005} for instance) by $p'=-[\rho+\Sigma\delta(r-R)+p]C'/2C+\delta(r-R)2\Pi/r$. Let us stress that these surface properties arise naturally as a consequence of matching the interior geometry with the exterior geometry at the surface in which the pressure of the relativistic star vanishes.


\noindent
\textsl{Summary.--}The effects of quantum vacuum polarization, which are widely expected to have a physical reality but to be negligible in most situations, are important enough to allow new static configurations of relativistic stars. We have shown this explicitly by constructing and solving a closed system of equations of stellar equilibrium (and in particular a generalized TOV equation) that includes semiclassical effects as described by quantum field theory in curved spacetimes and semiclassical gravity. The classical equations of stellar equilibrium are recovered at low density, but at high density, semiclassical effects cannot be disregarded. The analogy with the physics of neutron stars is appealing, with vacuum polarization providing a new kind of pressure of quantum-mechanical origin (analogous to the degeneracy pressure).

The exact solutions we have found describe relativistic stars with very specific properties. Although a detailed analysis will be presented elsewhere, let us give a brief summary here. Part of these properties correspond to a particular kind of horizonless ultra-compact object introduced \cite{Barcelo2007,Barcelo2009} and studied \cite{Barcelo2010,Barcelo2014e,Barcelo2014,Barcelo2015,Barcelo2016} during the last decade: black stars. The three main proposed characteristics of black stars, all of them satisfied by the exact solutions found here, are \cite{Barcelo2007,Barcelo2010} (i) an interior made of extremely dense matter supported by quantum vacuum polarization, (ii) a matter density profile proportional to the inverse of the square of the distance to the center, and (iii) a compactness $2Gm(r)/r$ arbitrarily close to (but slightly smaller than) unity for any value of $r$ (not only at the surface). On the other hand, other properties are strongly reminiscent of the gravastar proposal \cite{Mazur2004,Mottola2010,Mazur2015} (see also the generalizations in \cite{Visser2003,Lobo2005,Danielsson2017}). These are (iv) an equation of state of the form $p(r)\simeq-\rho(r)$ ---see \cite{Gliner1966,Sakharov1966} for arguments connecting this equation of state with hypothetical new states of matter at high densities--- (let us note that in the original gravastar model there is no classical matter in the interior, which is therefore optically thin, leading to specific observational imprints due to the defocusing of light rays \cite{Mazur2015} that should not be present in the case analyzed here), and (v) non-zero surface tension and energy density. That all these properties (i)-(v) arise naturally and together is remarkable. These results provide also an important theoretical basis for the study of phenomenological implications of horizonless ultra-compact alternatives to black holes, such as gravitational wave echoes \cite{Cardoso2016a,Cardoso2016,Abedi2016,Barcelo2017,Price2017,Nakano2017,Volkel2017,Mark2017,Maselli2017}.

\begin{acknowledgments}

Financial support was provided by the Claude Leon Foundation of South Africa. I would like to thank Carlos Barcel{\'o}, Luis J. Garay, Vitor Cardoso and Paolo Pani for useful discussions and for their critical reading of a previous version of the manuscript. I also appreciate the positive criticism received from anonymous referees, which has facilitated improving the manuscript.

\end{acknowledgments}

\bibliography{black_stars}	

\end{document}